\documentclass[prl, twocolumn, showpacs, nofootinbib, amsmath,amssymb, floatfix, eqsecnum]{revtex4-1}
\pdfoutput=1
\usepackage{amsmath}
\usepackage{braket}
\usepackage{amssymb}
\usepackage{amsthm}
\usepackage{algpseudocode}
\usepackage{algorithm}
\usepackage{amsfonts}
\usepackage{comment}
\usepackage[normalem]{ulem}
\usepackage{graphicx}
\usepackage{color,framed}
\usepackage{hyperref}
\usepackage{times}
\usepackage{enumerate}
\usepackage{lipsum}
\usepackage{slashed}
\usepackage{url}
\usepackage{bbm}
\usepackage{chngcntr}
\counterwithout{equation}{section}
\usepackage{tikz,pgfplots}
\usepackage{pgfplotstable}
\usepgfplotslibrary{fillbetween}

\hypersetup{
    colorlinks=true, 
    linktoc=all,     
    linkcolor=blue,  
}

\def \beq {\begin{equation}}
\def \eeq {\end{equation}}
\def \beqa {\begin{eqnarray}}
\def \eeqa {\end{eqnarray}}
\def \bseq {\begin{subequations}}
\def \eseq {\end{subequations}}

\newcommand{\sign}{{\rm sgn}\,}

\newcommand{\D}[1]{\text{d}#1}

\pgfplotsset{compat=1.18}
\begin{document}

\title{Continuous Hamiltonian dynamics on digital quantum computers without discretization error}

\author{Etienne Granet and Henrik Dreyer}
\affiliation{Quantinuum, Leopoldstrasse 180, 80804 Munich, Germany}
\date{\today}

\begin{abstract}
We introduce an algorithm to compute Hamiltonian dynamics on digital quantum computers that requires only a finite circuit depth to reach an arbitrary precision, i.e. achieves zero discretization error with finite depth. This finite number of gates comes at the cost of an attenuation of the measured expectation value by a known amplitude, requiring more shots per circuit. The gate count for simulation up to time $t$ is $\mathcal{O}(t^2\mu^2)$ with $\mu$ the $1$-norm of the Hamiltonian, without dependence on the precision desired on the result, providing a significant improvement over previous algorithms. The only dependence in the norm makes it particularly adapted to non-sparse Hamiltonians. The algorithm generalizes to time-dependent Hamiltonians, appearing for example in adiabatic state preparation. These properties make it particularly suitable for present-day relatively noisy hardware that supports only circuits with moderate depth. 
\end{abstract}

\maketitle

\textbf{\emph{Introduction.}}--- Hamiltonian dynamics is one of the most promising application of current and near-term quantum computers \cite{feynman2018simulating,lloyd1996universal}. It is believed to be exponentially faster to perform on quantum computers and finds multiple applications, either directly or indirectly as subroutines of other quantum algorithms \cite{abrams1999quantum,harrow2009quantum,farhi2000quantum, yang2023prethermalisation, lu2021finiteenergies, schuckert2022probing}. Given a Hamiltonian $H$ and a simulation time $t$, there exist several ways to implement $U(t)=e^{itH}$ on a digital quantum computer where only one or two-qubit gates can be used. The simplest approach uses product formulas such as Trotterization to decompose the dynamics into elementary gates $U(t)\approx e^{itH_1}...e^{itH_n}$ \cite{lloyd1996universal,suzuki1990fractal,suzuki1991general}. These methods have been generalized and improved in many ways \cite{berry2007efficient,childs2019nearly,childs2021theory,low2023complexity,zeng2022simple,cho2022doubling,tomesh2021optimized,aharonov2003adiabatic,childs2012hamiltonian,haah2021quantum,leadbeater2023non,fitzpatrick2021evaluating,ostmeyer2022optimised,mc2023classically}. In particular, the introduction of randomness  \cite{campbell2017shorter,childs2019faster,faehrmann2022randomizing,wan2022randomized,wang2023qubit,gong2023improved} such as in the qDRIFT algorithm \cite{campbell2019random,ouyang2020compilation,chen2021concentration,nakaji2023qswift,kiss2023importance,hagan2022composite,huggins2021virtual,berry2020time,pocrnic2023composite} has particularly good theoretical performance for non-sparse Hamiltonians $H$, i.e. when the number of terms in the Hamiltonian grows faster than $\mathcal{O}(L)$ where $L$ is the number of qubits. These algorithms have in common that they only \emph{approximate} the continuous time evolution with finite-depth circuits, and display discretization errors (sometimes called \emph{Trotter errors}) that vanish only with infinitely many gates. This is particularly problematic for adiabatic state preparation (since Trotter errors generically lead to an effective heating) \cite{aspuru2005simulated,farhi2000quantum,albash2018adiabatic} and in applications like quantum chemistry \cite{su2021fault,reiher2017elucidating,bauer2020quantum,mcardle2020quantum,motta2022emerging,kim2022fault} where extreme precision is desired. More sophisticated algorithms have better theoretical scalings, such as linear combination of unitaries \cite{childs2012hamiltonian,berry2015simulating,berry2014exponential} (LCU), quantum signal processing \cite{low2017optimal,low2019hamiltonian,low2018hamiltonian,low2016methodology,kikuchi2023realization} or quantum walks \cite{berry2009black,childs2010relationship}, but require a large resource overhead. As a consequence, at least in the near term, exact continuous time dynamics is considered to be the prerogative of non-gate-based, analog quantum simulators \cite{buluta2009quantum,blatt2012quantum,aspuru2012photonic,bloch2012quantum,georgescu2014quantum}.

In this Letter we introduce an algorithm to compute Hamiltonian dynamics of observables $\langle \mathcal{M}(t)\rangle=\langle 0|e^{iHt}\mathcal{M}e^{-iHt}|0\rangle$ on digital quantum computers without discretization error even with a finite number of gates, for any Hamiltonian $H$ and unitary $\mathcal{M}$. The expectation value $\langle \mathcal{M}(t)\rangle$ is obtained as the average over random circuits drawn from a judiciously chosen distribution, multiplied by a known factor. One can choose freely the angle $\tau$ of the rotations entering the circuit, only modifying the multiplicative factor and not the precision of the result. It allows one to decrease the gate count at the cost of an increased number of shots on a quantum computer, yielding a natural noise-mitigation technique. The optimal choice of gate angle yields a shot overhead of order $\mathcal{O}(1)$ with a gate count per circuit $\mathcal{O}(t^2\mu^2)$ where $\mu$ is the $1$-norm of the Hamiltonian, \emph{without any} dependence on the precision desired, contrary to all previous algorithms. Since the gate count depends only on $\mu$, the algorithm is particularly adapted to non-sparse Hamiltonians. It also straightforwardly generalizes to time-dependent Hamiltonians without any overhead. We demonstrate the algorithm with numerical simulations of the electronic structure of the stretched water molecule and of a 2D Ising model where it outperforms both Trotter formulas as well as other randomised compilation techniques.

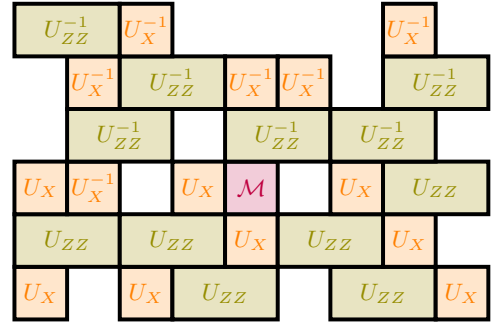
\begin{figure}
\begin{center}
\begin{tikzpicture}[scale=0.7]

\draw[black, line width=1.5pt,fill=orange!20] (0, 1) rectangle (1, 1+1);
\draw node[orange] at (0.5,1+0.5) { $U_X$};

\draw[black, line width=1.5pt,fill=orange!20] (2+0, 1) rectangle (2+1, 1+1);
\draw node[orange] at (2+0.5,1+0.5) { $U_X$};

\draw[black, line width=1.5pt,fill=olive!20] (3+0, 1+0) rectangle (3+2, 1+1);
\draw node[olive] at (3+1,1+0.5) { $U_{ZZ}$};

\draw[black, line width=1.5pt,fill=olive!20] (6+0, 1+0) rectangle (6+2, 1+1);
\draw node[olive] at (6+1,1+0.5) { $U_{ZZ}$};

\draw[black, line width=1.5pt,fill=olive!20] (0+0, 2+0) rectangle (0+2, 2+1);
\draw node[olive] at (0+1,2+0.5) { $U_{ZZ}$};

\draw[black, line width=1.5pt,fill=olive!20] (2+0, 2+0) rectangle (2+2, 2+1);
\draw node[olive] at (2+1,2+0.5) { $U_{ZZ}$};

\draw[black, line width=1.5pt,fill=orange!20] (4+0, 2+0) rectangle (4+1, 2+1);
\draw node[orange] at (4+0.5,2+0.5) { $U_X$};

\draw[black, line width=1.5pt,fill=orange!20] (7+0, 2+0) rectangle (7+1, 2+1);
\draw node[orange] at (7+0.5,2+0.5) { $U_X$};

\draw[black, line width=1.5pt,fill=olive!20] (5+0, 2+0) rectangle (5+2, 2+1);
\draw node[olive] at (5+1,2+0.5) { $U_{ZZ}$};

\draw[black, line width=1.5pt,fill=orange!20] (0+0, 3+0) rectangle (0+1, 3+1);
\draw node[orange] at (0+0.5,3+0.5) { $U_X$};

\draw[black, line width=1.5pt,fill=orange!20] (1+0, 3+0) rectangle (1+1, 3+1);
\draw node[orange] at (1+0.5,3+0.5) { $U_X^{-1}$};

\draw[black, line width=1.5pt,fill=orange!20] (3+0, 3+0) rectangle (3+1, 3+1);
\draw node[orange] at (3+0.5,3+0.5) { $U_X$};

\draw[black, line width=1.5pt,fill=orange!20] (6+0, 3+0) rectangle (6+1, 3+1);
\draw node[orange] at (6+0.5,3+0.5) { $U_X$};

\draw[black, line width=1.5pt,fill=olive!20] (1+0, 4+0) rectangle (1+2, 4+1);
\draw node[olive] at (1+1,4+0.5) { $U_{ZZ}^{-1}$};

\draw[black, line width=1.5pt,fill=orange!20] (8+0, 1+0) rectangle (8+1, 1+1);
\draw node[orange] at (8+0.5,1+0.5) { $U_X$};

\draw[black, line width=1.5pt,fill=olive!20] (7+0, 3+0) rectangle (7+2, 3+1);
\draw node[olive] at (7+1,3+0.5) { $U_{ZZ}$};

\draw[black, line width=1.5pt,fill=purple!20] (4+0, 3+0) rectangle (4+1, 3+1);
\draw node[purple] at (4+0.5,3+0.5) { $\mathcal{M}$};

\draw[black, line width=1.5pt,fill=olive!20] (4+0, 4+0) rectangle (4+2, 4+1);
\draw node[olive] at (4+1,4+0.5) { $U_{ZZ}^{-1}$};

\draw[black, line width=1.5pt,fill=olive!20] (6+0, 4+0) rectangle (6+2, 4+1);
\draw node[olive] at (6+1,4+0.5) { $U_{ZZ}^{-1}$};

\draw[black, line width=1.5pt,fill=orange!20] (1+0, 5+0) rectangle (1+1, 5+1);
\draw node[orange] at (1+0.5,5+0.5) { $U_X^{-1}$};

\draw[black, line width=1.5pt,fill=orange!20] (4+0, 5+0) rectangle (4+1, 5+1);
\draw node[orange] at (4+0.5,5+0.5) { $U_X^{-1}$};

\draw[black, line width=1.5pt,fill=orange!20] (5+0, 5+0) rectangle (5+1, 5+1);
\draw node[orange] at (5+0.5,5+0.5) { $U_X^{-1}$};

\draw[black, line width=1.5pt,fill=olive!20] (2+0, 5+0) rectangle (2+2, 5+1);
\draw node[olive] at (2+1,5+0.5) { $U_{ZZ}^{-1}$};

\draw[black, line width=1.5pt,fill=olive!20] (7+0, 5+0) rectangle (7+2, 5+1);
\draw node[olive] at (7+1,5+0.5) { $U_{ZZ}^{-1}$};

\draw[black, line width=1.5pt,fill=olive!20] (0+0, 6+0) rectangle (0+2, 6+1);
\draw node[olive] at (0+1,6+0.5) { $U_{ZZ}^{-1}$};

\draw[black, line width=1.5pt,fill=orange!20] (2+0, 6+0) rectangle (2+1, 6+1);
\draw node[orange] at (2+0.5,6+0.5) { $U_X^{-1}$};

\draw[black, line width=1.5pt,fill=orange!20] (7+0, 6+0) rectangle (7+1, 6+1);
\draw node[orange] at (7+0.5,6+0.5) { $U_X^{-1}$};

\end{tikzpicture}
\caption{Example of a configuration for a transverse-field Ising Hamiltonian, with $U_O=e^{i\tau O}$ acting on qubits at the bottom of the drawing. The average over such configurations converges to the exact time-evolved expectation value $\langle 0 | e^{iHt} \mathcal{M} e^{-iHt}|0\rangle$, up to a known multiplicative constant.}
\label {tetris}
\end{center}
\end {figure}

\textbf{\emph{Simulating small-angle gates with large-angle gates.}}---We first introduce the basic mechanism that underpins our algorithm. For a real number $0\leq p\leq 1$, an angle $\tau$ and $O$ any operator such that $O^2=I$, we have the equality
\begin{equation}
\label{eq_basic_mechanism}
    1-p+pe^{i\tau O}=\lambda e^{i\tau' O}\,,
\end{equation}
where
\begin{equation}
    \tan\tau '=\frac{p\sin\tau}{1-p+p\cos \tau}\,,\qquad \lambda=\frac{p\sin \tau}{\sin \tau'}\,.
\end{equation}
This relation can be used to realize an effective angle $0<\tau'<\tau$ by applying $e^{i\tau O}$ with probability
\begin{equation}
    p=\frac{\tan \tau'}{\sin\tau+(1-\cos\tau)\tan\tau'}\,.
\end{equation}
To see this, we consider a circuit that contains $G$ many rotations $e^{i\tau' O}$ and that produces the wave function $|\psi\rangle$. For a subset $S$ of these rotations, we denote $|\psi_S\rangle$ the wave function obtained with the same circuit but with deleting the rotations $e^{i\tau' O}$ contained in $S$ and replacing the remaining rotations $e^{i\tau' O}$ not in $S$ by $e^{i\tau O}$. Then, for any observable $\mathcal{M}$, by repeatedly applying~(\ref{eq_basic_mechanism}), we have
\begin{equation}
   \sum_{S,S'} (1-p)^{|S|+|S'|}p^{2G-|S|-|S'|} \langle \psi_{S'}|\mathcal{M}|\psi_S\rangle=\lambda^{2G} \langle \psi|\mathcal{M}|\psi\rangle\,,
\end{equation}
where the sum runs over all the possible subsets $S,S'$ of gates to delete in the original circuit. The expectation value $\langle \psi|\mathcal{M}|\psi\rangle$ that originally involves a circuit with angle $\tau'>0$ can thus be exactly computed with a circuit involving only larger angles $\tau>\tau'$ but in which some rotations are randomly removed. This however comes with an attenuation factor $\lambda^{2G}<1$  which increases the total number of shots required to reach a given precision on $\langle \psi|\mathcal{M}|\psi\rangle$. The same reasoning applies to rotations with a negative angle $\tau'<0$, with then $\tau<\tau'$. We note that similar mechanisms have been used recently to implement arbitrary rotation angles \cite{koczor2023probabilistic}.

\textbf{\emph{Continuous time limit.}}---We now consider a Hamiltonian $H$ that we write as a linear combination of $N$ operators $O_n$ that satisfy $O_n^2=I$, i.e.
\begin{equation}\label{ham}
    H=\sum_{n=1}^N c_n O_n\,,
\end{equation}
which is always possible by writing it as a sum of product of Pauli matrices. We would like to compute the expectation value of a unitary $\mathcal{M}$ within the wave function $|\psi(t)\rangle=e^{iHt}|\psi(0)\rangle$. According to the Trotter-Suzuki formula, we have
\begin{equation}\label{trotter}
    e^{itH}=\underset{\tau'\to 0^+}{\lim} (e^{i\tau' c_1 O_1}...e^{i\tau' c_N O_N})^{t/\tau'}\,.
\end{equation}
For each $n=1,...,N$, we choose $0<\tau_n\leq \pi/2$ and implement $e^{i\tau' c_n O_n}$ using the previously explained protocol with angle $\tau_n$. In the sequence of rotations appearing on the right-hand side of \eqref{trotter}, instead of each rotation $e^{i\tau' c_n O_n}$ we thus apply $e^{i\tau_n \sign(c_n)O_n}$ with probability $p_n$ given by
\begin{equation}
    p_n=\frac{\tau' |c_n|}{\sin\tau_n }+\mathcal{O}((\tau')^2)\,.
\end{equation}
The corresponding attenuation factor $\lambda_n$ is
\begin{equation}
    \lambda_n=1-\tau' |c_n| \tan(\tau_n/2)+\mathcal{O}((\tau')^2)\,.
\end{equation}
In the limit $\tau'\to 0$, the probability of picking a rotation $p_n\to 0$ becomes a rare event and we converge to a Poisson process. Namely, for each rotation $e^{i\tau_n \sign(c_n)O_n}$ we obtain a sequence of gate times $0<t_n^{(1)}<...<t_n^{(m_n)}<t$ drawn from a Poisson process with rate $|c_n|/\sin \tau_n$. For a given realization $T$ of all these times for different $n$'s, we denote $|\psi_T\rangle$ the wave function obtained by applying the rotations $e^{i\tau_n \sign(c_n)O_n}$ ordered by time of occurrence $t_n^{(i)}$, irrespective of $n$. Such a configuration $|\psi_T\rangle$ is depicted in Fig \ref{tetris}. Then, denoting $\mathbb{E}$ the expectation value with respect to two independent configurations $T,T'$, we have
\begin{equation}\label{formula}
   \langle \psi(t)|\mathcal{M}|\psi(t)\rangle=\frac{\mathbb{E}[\langle \psi_{T'}|\mathcal{M}|\psi_T\rangle]}{\lambda_{\rm att}}\,,
\end{equation}
with the total attenuation
\begin{equation}\label{attenuation}
    \lambda_{\rm att}=\exp\left(-2t \sum_{n=1}^N |c_n|\tan(\tau_n/2) \right)\,.
\end{equation}
This is the fundamental relation that defines our algorithm. Remarkably, the exact expectation value at time $t$ under continuous time dynamics can be obtained \emph{without} discretization error, by only applying rotations with application time $\tau_n>0$ fixed. Moreover, the average number of rotations in each circuit is
\begin{equation}
    N_{\rm rot}=2t\sum_{n=1}^N \frac{|c_n|}{\sin \tau_n}\,,
\end{equation}
which is finite and only controlled by $\tau_n$ and not by the precision wanted on the result. We note that the statistical variance in estimating $\langle\psi(t)|\mathcal{M}|\psi(t)\rangle$ is bounded by $\lambda_{\rm att}^{-1}$. An important consequence of this attenuation factor $\lambda_{\rm att}$ is thus to increase the number of shots required by a factor $\lambda_{\rm att}^{-2}$ to reach a given precision.

\textbf{\emph{Statement of the algorithm.}}--- We state here the algorithm deduced from the previous explanations. It takes as input a Hamiltonian $H$ decomposed as in \eqref{ham}, a unitary observable $\mathcal{M}$, a time $t$ and an initial state $|\psi(0)\rangle$, and outputs $ \langle \psi(t)|\mathcal{M}|\psi(t)\rangle$. It takes $0<\tau_n\leq \pi/2$ as $N$ parameters.

\begin{enumerate}
    \item Prepare the system in the initial state $|\psi(0)\rangle$.
    \item For each $n\in\{1,...,N\}$, draw an integer $m_n$ from a Poisson distribution with parameter $\frac{|c_n|t}{\sin |\tau_n|}$. Then draw $m_n$ real numbers $t_n^{(i)}$, where $i\in\{1,...,m_n\}$, uniformly at random between $0$ and $t$. These $M\equiv \sum_{n=1}^N m_n$ numbers are collected into a set $T$. 
    
    \item Deduce then the sequence $k_1,...,k_M\in\{1,...,N\}$ such that the index $n$ corresponding to the $j$-th smallest element of $T$ is $k_j$. For $m=1,...,M$ in this order, apply the rotation $e^{i\tau_{k_m} \sign(c_{k_m})O_{k_m}}$ on the system.
    \item Apply the observable $\mathcal{M}$ on the system.
    \item Repeat steps (2) and (3) with $\tau_n$ replaced by $-\tau_n$.
    \item Measure the overlap with the initial state and divide the result by $\lambda_{\rm att}$ given in \eqref{attenuation}.
    On a quantum computer, this step requires to perform a Hadamard test.
    \item Repeat steps (1) to (6) a number of times and average the results.
    
\end{enumerate}
 When measuring the overlap with the initial state, one has to choose a number of shots $S$ to do per configuration, i.e. how to distribute the resources between number of configurations and accuracy of each configuration. If we neglect compilation time, the optimal number of shots can be seen to be $S=1$ \cite{supp}.

 \textbf{\emph{Measurement of the time evolution operator.}}--- The algorithm presented above can be adapted to computing the amplitude $\langle \psi(0)|e^{iHt}|\psi(0)\rangle$. It is often called Loschmidt amplitude in the condensed matter literature and can be used to determine ground state energies as well as microcanical expectation values \cite{lu2021finiteenergies,hemery2023measuring}. With the same notations as in Eq.~\eqref{formula}, we have $\langle \psi(0)|e^{iHt}|\psi(0)\rangle=\mathbb{E}[\langle \psi(0)|\psi_T\rangle]/\sqrt{\lambda_{\rm att}}$. One thus skips steps (4) and (5) in the algorithm above, and replace the attenuation factor $\lambda_{\rm att}$ by $\sqrt{\lambda_{\rm att}}$ in step (6).

\textbf{\emph{Optimal angle on a perfect quantum computer.}}--- Formula \eqref{formula} holds for any choice of gate angle $\tau_n$. Increasing the gate angle $\tau_n$ decreases the number of rotations per circuit $N_{\rm rot}$, but also decreases the attenuation factor $\lambda_{\rm att}$, and so increases the number of shots $\lambda_{\rm att}^{-2}$ that have to be performed to reach a given precision. The total number of rotations $R(\{\tau_n\})=N_{\rm rot}\lambda_{\rm att}^{-2}$ to perform to get a precision of order $\mathcal{O}(1)$ on the result is thus a non-trivial function of the gate angles $\tau_n$, that reaches a minimum at some $\tau_n^*$ that can be computed numerically. At large $t$, we find analytically that the optimal angle is
\begin{equation}\label{optimaltau}
    \tau^*_n=\frac{1}{2t\mu}\,,
\end{equation}
for all $n=1,...,N$, where $\mu=\sum_{n=1}^N|c_n|$ is the $1$-norm of the Hamiltonian \cite{supp}. In that limit we find $\lambda_{\rm att}=e^{-1/2}$ which remains finite and $N_{\rm rot}=4t^2\mu^2$. Hence the gate count per circuit scales as
\begin{equation}
    \mathcal{O}(t^2\mu^2)\,.
\end{equation}
The number of shots to perform to reach precision $\epsilon$ scales as $\mathcal{O}(\epsilon^{-2})$. But remarkably, the precision $\epsilon$ \emph{does not} enter the gate count, which remains finite even when $\epsilon\to 0$. This differs from usual algorithms such as Trotter, qDRIFT or LCU, whose gate count we recall in Fig \ref{tabular}.

\begin{figure}[H]
\begin{center}
\begin{tabular}{|c|c|c|c|}
\hline
      Trotter $K$-th order & qDRIFT & LCU & our algorithm \\
    \hline
      $(Nt)^{1+1/K}\epsilon^{-1/K}$ & $t^2\mu^2\epsilon^{-1}$ & $N\mu t \tfrac{\log(\mu t/\epsilon)}{\log\log \mu t/\epsilon}$ & $t^2\mu^2$\\
    \hline
    
\end{tabular}
\end{center}
\caption{Gate count of different algorithms for Hamiltonian dynamics. $N$ refers to the number of terms in the Hamiltonian, $\mu$ its $1$-norm, $t$ the simulation time and $\epsilon$ the precision reached on the result.}
\label {tabular}
\end{figure}

\textbf{\emph{Optimal angle on a noisy quantum computer.}}---
The optimal angle value \eqref{optimaltau} holds for a perfect, noiseless quantum computer. Let us explain how this formula is modified in presence of noise. For simplicity, we assume that the application of each gate $e^{i\tau_n \sign(c_n)O_n}$ comes with a signal attenuation $e^{-r_n}$ with some $r_n>0$. This kind of noise model is known to be a rough first approximation of the effect of the noise on current hardware. Moreover, other types of noise can be converted into this global depolarizing noise through some noise mitigation techniques \cite{yang2023prethermalisation,temme2017error,self2022protecting}. In state-of-the-art devices typical two-qubit gate fidelities are $e^{-r} \sim 99.8\%$, in which case $r \sim 2\mathrm{e}{-3}$ \cite{moses2023racetrack}. There are on average $\frac{t|c_n|}{\sin\tau_n}$ such rotations per configuration. The damping due to hardware imperfection is thus
\begin{equation}
    q_{\rm att}=\exp\left(-2t\sum_{n=1}^N \frac{r_n|c_n|}{\sin\tau_n} \right)\,.
\end{equation}
The total attenuation of the signal is $\lambda_{\rm att} q_{\rm att}$. Assuming $r_n$ small, the optimal times $\tau_n^*$ that minimize the total gate count at large $t$ are now \cite{supp}
\begin{equation}\label{optimal}
    \tau_n^*=\sqrt{2r_n}\,.
\end{equation}

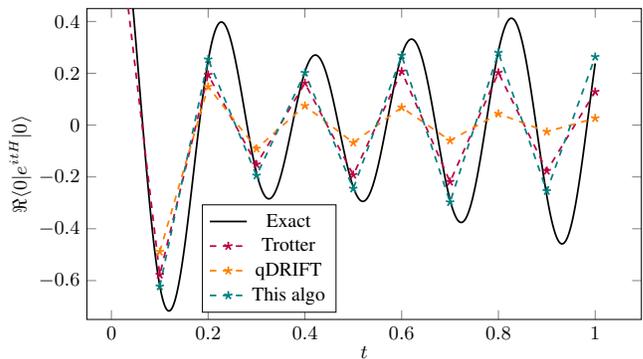
\begin{figure}[tb]
\begin{center}
\begin{tikzpicture}[scale=0.82]
\begin{axis}[ xlabel=$t$, ylabel=$\Re \langle 0| e^{itH}|0\rangle$,scale=0.55,ylabel style={yshift=-0.1cm},xlabel style={yshift=0.1cm}, ymax=0.45,width=1\textwidth,height=0.6\textwidth,legend style={at={(0.45,0.37)}},ymin=-0.75],

\addplot[color=black,smooth,thick] table {plots_N12_h2_exact.txt};
\addplot[color=purple,mark=star,dashed,thick] table {plots_N12_h2_trotter.txt};
\addplot[color=orange,mark=star,dashed,thick] table {plots_N12_h2_qdrift.txt};
\addplot[color=teal,mark=star,dashed,thick] table {plots_N12_h2_tetris.txt};

\addlegendentry{Exact}
\addlegendentry{Trotter}
\addlegendentry{qDRIFT}
\addlegendentry{This algo}

 \end{axis} 
\end{tikzpicture}
\caption{Real part of the Loschmidt amplitude $\langle 0| e^{itH}|0\rangle$ in the 2D Ising model at $h=2$ in size $3\times 4$ with $|0\rangle$ the $Z=+1$ product state, comparing Trotter, qDRIFT and our algorithm, all with the same gate angle and Trotter step equal to $\tau=0.1$, and all with $10^5$ shots per point. For qDRIFT and our algorithm, the $10^5$ shots are distributed over $10^3$ random circuits with $10^2$ shots per circuit. 
}
\label {exact}
\end{center}
\end {figure}

\textbf{\emph{Time-dependent Hamiltonian.}}--- The algorithm can be generalized straightforwardly to dynamics with a time-dependent Hamiltonian $H(t)$. Decomposing
\begin{equation}
    H(t)=\sum_{n=1}^N c_n(t) O_n\,,
\end{equation}
with time-dependent coefficients $c_n(t)$, and denoting $|\psi(t)\rangle=\mathcal{T}\exp\left(i\int_0^t\D{s} H(s)\right)|\psi(0)\rangle$, the expectation value of operators $\mathcal{M}$ can again be written as an average over configurations \eqref{formula}. The configurations $|\psi_T\rangle$ are produced by drawing times $0<t_n^{(1)}<...<t_n^{(m_n)}<t$ from a Poisson process with time-dependent rate $|c_n(s)|/\sin \tau_n$, for each $n=1,...,N$, and applying on the initial wave function each gate $e^{i\tau_n \sign(c_n(s))O_n}$ ordered by time of occurrence $s=t_n^{(i)}$. The attenuation factor is then obtained by replacing $t |c_n|$ by $\int_0^t \D{s} |c_n(s)|$.

In practice, we proceed as follows. We introduce the function
\begin{equation}
    z_n(u)=\int_0^u\D{s} |c_n(s)|\,,
\end{equation}
as well as $z_n^{-1}(u)$ its reciprocal, i.e. such that $z_n^{-1}(z_n(u))=u$. We replace step (2) of the time-independent case by
\begin{enumerate}
\setcounter{enumi}{1}
    \item For each $n\in\{1,...,N\}$, draw an integer $m_n$ from a Poisson distribution with parameter $\frac{z_n(t)}{\sin |\tau_n|}$. Then draw $m_n$ real numbers $\tilde{t}_n^{(i)}$, where $i\in\{1,...,m_n\}$, uniformly at random between $0$ and $z_n(t)$, and set $t_n^{(i)}=z_n^{-1}(\tilde{t}_n^{(i)})$. These $M\equiv \sum_{n=1}^N m_n$ numbers $t_n^{(i)}$ are collected into a set $T$. 
\end{enumerate}
Then in step (3), the rotation applied is $e^{i\tau_{k_m} \sign(c_{k_m}(s))O_{k_m}}$ where $s\in \{t_n^{(i)}\}$ is the time at which the rotation is applied. For the backward propagation one has to reverse the order of the gates. The total attenuation $\lambda_{\rm att}$ of step (6) is
\begin{equation}
    \lambda_{\rm att}=\exp\left(-2\sum_{n=1}^N z_n(t) \tan(\tau_n/2) \right)\,.
\end{equation}
This algorithm produces the \emph{exact} time evolution of the wave function under the time-dependent Hamiltonian $H(t)$, without errors arising from discretizing  the continuous function $c_n(s)$ or Trotter errors.

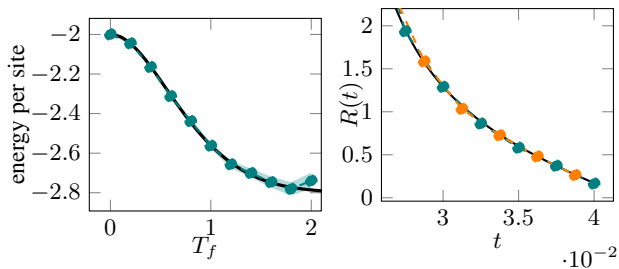
\begin{figure}[tb]
\begin{center}
\begin{tikzpicture}[scale=1.]
\begin{axis}[ xlabel=$T_f$, ylabel=energy per site,scale=0.45,xlabel style={yshift=0.19cm},ylabel style={yshift=-0.2cm},xmax=2.1]

\addplot[color=teal!30,no marks,name path=F] table[y expr={\thisrowno{1}+\thisrowno{2}}]{adiabatic_T_tetris_variance_100000.txt};
\addplot[color=teal!30,no marks,name path=G] table[y expr={\thisrowno{1}-\thisrowno{2}}]{adiabatic_T_tetris_variance_100000.txt};
\addplot[color=teal!30]fill between[of=F and G];

\addplot[color=black,smooth,very thick] table {adiabatic_T_exact.txt};

\addplot[color=teal,thick,dashed,mark=*] table {adiabatic_T_tetris_variance_100000.txt};

 \end{axis} 
\end{tikzpicture}
\begin{tikzpicture}[scale=1.0]
\begin{axis}[ xlabel=$t$, ylabel=$R(t)$,scale=0.45,ymax=2.2,xlabel style={yshift=0.1cm},ylabel style={yshift=-0.5cm}]

\addplot[color=black,thick,smooth] table {h2o_theory2.txt};
\addplot[color=teal,thick,dashed,mark=*] table {h2o_noiseless2.txt};
\addplot[color=orange,thick,dashed,mark=*] table {h2o_noisy2.txt};

 \end{axis} 
\end{tikzpicture}

\caption{Left: adiabatic state preparation (see text), noiseless simulations with $10^5$ circuits and $\tau=0.04$ (teal, the shade indicating one standard deviation), and exact value (black). Right: chemistry Hamiltonian (see text), with gate angles \eqref{optimal}, with $2000$ circuits with $20$ shots each.}

\label{adiabatic}
\end{center}
\end {figure}

\textbf{\emph{Background Hamiltonian.}}--- The algorithm has the following simple improvement. Let us consider a subset of gates $\mathcal{O}\subset \{O_n\}_{n=1,..,N}$ that all commute among themselves, i.e. $[O,O']=0$ for all $O,O'\in\mathcal{O}$, and denote $H_{\rm background}=\sum_{O_p\in\mathcal{O}}c_p O_p$. If for these terms we take the limit of zero angle $\tau_p\to 0$ in the algorithm, the Poisson process almost always picks rotations of terms in $\mathcal{O}$ with very short angle. But since they commute this can be implemented with a finite product of $e^{i\tau O}$ with $\tau$ obtained by summing the different angles. Hence the algorithm is modified as follows. In step (2) we only draw times corresponding to terms \emph{not} in $\mathcal{O}$, and use them to produce an ordered sequence of gates as in step (3). However, between each application of these rotations $e^{i\tau_n O_n}$ and $e^{i\tau_m O_m}$ at times $s_n<s_m$, we apply $e^{i \delta s H_{\rm background}}$ with $\delta s=s_m-s_n$ the difference between the two instants at which the rotations are applied. This reduces the attenuation factor $\lambda_{\rm att}$, which is now computed \emph{only} including the terms \emph{not} in $\mathcal{O}$. The Hamiltonian $H_{\rm background}$ thus plays the role of a ``background" Hamiltonian that is always applied on the system, but on top of which rotations $e^{i\tau_n O_n}$ with $O_n\notin \mathcal{O}$ are applied.

\textbf{\emph{Numerical tests.}}---
We now show numerical simulations of our algorithm. We start with a noisy implementation  on a non-sparse Hamiltonian as routinely encountered in quantum chemistry. Hamiltonians describing molecular electronic structure problems are typically decomposed in a basis set of fermionic orbitals, where they read
\begin{equation}
    H=\sum_{ij=1}^N h_{ij}c^\dagger_i c_j+\sum_{ijkl=1}^N h_{ijkl} c^\dagger_i c^\dagger_j c_k c_l\,,
\end{equation}
with $c$'s canonical fermionic operators and $N$ the number of orbitals considered. Written in terms of Pauli gates through a Jordan-Wigner transformation, these molecular Hamiltonians involve thus $\mathcal{O}(N^4)$ terms with each $\mathcal{O}(N)$ Pauli gates, incurring a prohibitively large gate count for any Trotter implementation of the dynamics. We consider a H${}_2$O water molecule in a STO-6G basis with $14$ orbitals, and use our algorithm to study the Loschmidt amplitude
\begin{equation}
    \mathcal{L}(t)=\langle HF|e^{itH}|HF\rangle\,,
\end{equation}
where $|HF\rangle$ denotes the Hartree-Fock ground state. We take a geometry with an angle H-O-H of $105^o$ and with an elongated distance H-O of $2.2$ \r{A}, to have a significant difference between the exact ground state and $|HF\rangle$. On a noisy hardware, we propose to compute the ratio
\begin{equation}\label{Rt}
    R(t)=\frac{\Im \mathcal{L}(t)}{\Re \mathcal{L}(t)}\,.
\end{equation}
This ratio is not sensitive to the depolarizing part of the hardware noise, but still contains information about the energy levels in the oscillation frequencies. In the right panel of Figure \ref{adiabatic} we show a noisy simulation of our algorithm that includes a depolarizing channel with probability $0.002$ after each $2$-qubit gate, and observe very good agreement.

Next, we show numerical results for the time-dependent case of our algorithm. For this case, we consider the 2D square lattice Ising model in a transverse field $h$, whose Hamiltonian is
\begin{equation}\label{ising}
    H=-\sum_{\langle i,j\rangle}Z_iZ_j-h\sum_{j}X_j\,,
\end{equation}
with periodic boundary conditions. We implement the adiabatic preparation of the ground state at $h_f=2.5$, starting from $h=0$. For a final time $T_f$ we consider the time-dependent magnetic field $ h(t)=h_f \sin(\tfrac{\pi}{2}\sin(\tfrac{\pi t}{2T_f})^2)^2$~\cite{wei2023adiabatic}.
In the left panel of Figure \ref{adiabatic} we show the final energy per site obtained at $t=T_f$, as a function of $T_f$, together with the error bars for a fixed number of shots. We see the agreement with the exact values, as well as the broadening of the error bars as $T_f$ grows at fixed $\tau$, as imposed by \eqref{attenuation}.

Finally, we compare our algorithm to first-order Trotter and qDRIFT in the case of the 2D square lattice Ising model \eqref{ising} with magnetic field $h=2$, which is a typical model studied in condensed matter. In Fig \ref{exact} we plot the Loschmidt amplitude $\langle 0|e^{iHt}|0\rangle$ starting from the $Z=+1$ product state $|0\rangle$, using a same gate angle $\tau=0.1$ for our algorithm and for qDRIFT, and with a same time step $0.1$ for Trotter. We take $10^5$ shots per point (which is necessary to have a precision $<0.01$ with high confidence), and distribute them over $10^3$ random circuits with $10^2$ shots per circuit for our case and for qDRIFT. We see that our algorithm gives almost perfect results. In particular, it works considerably better than qDRIFT. We also see that it performs better than  Trotter. This shows the practical use of our algorithm.

\textbf{\emph{Interpolation between quantum time evolution and path integral representation.}}--- Some comments about the upper and lower limits of the angle $\tau$ are in order. When $\tau\to 0$, the different rotations $e^{i\tau O_n}$ commute at order $\tau$ and there are increasingly more rotations drawn from the Poisson process. With probability $1$ in this limit, a single configuration $|\psi_T\rangle$ is equal to the exact time evolution $e^{iHt}|0\rangle$, and the attenuation factor \eqref{attenuation} is equal to $1$. When $\tau=\pi/2$, we have $e^{i\tau O_n}=iO_n$, and so when $O_n$ is a string of Pauli operators, the rotations can be implemented with only one-qubit gates. Each configuration $|\psi_T\rangle$ remains thus a product state in the $Z$ basis and describes a classical trajectory in discrete time. Thus \eqref{formula} becomes akin to a path integral representation, in the sense that a quantum expectation value is expressed as an averaging over exponentially many classical trajectories. 

These two limits provide thus an interesting interpretation of our algorithm as an \emph{interpolation} between a single quantum trajectory for $\tau=0$, and exponentially many classical trajectories  for $\tau=\pi/2$. Intermediate values of $\tau$ near $\pi/2$ allow one to decrease the entanglement present in each configuration $|\psi_T\rangle$ compared to the exact time-evolved state $e^{iHt}|0\rangle$, but at the cost of increasing the number of samples required through the attenuation factor \eqref{attenuation}. This could find applications in the classical simulation of quantum dynamics.

\textbf{\emph{Discussion.}}--- We presented a quantum algorithm to compute continuous Hamiltonian dynamics that requires a circuit depth that is independent of the desired precision $\epsilon$, contrary to many algorithms that require an infinite depth when $\epsilon\to 0$. We show that by applying randomly gates on the initial state according to a well-chosen distribution, and multiplying the result by a known amplification factor, one achieves zero discretization error while having finite-depth circuits. This amplification factor however increases the number of shots required to reach a given precision. The algorithm is particularly adapted to non-sparse Hamiltonians as the gate count only depends on the $1$-norm and not the number of terms.

The algorithm and its mechanisms can be generalized in a number of ways. For example, decomposing a circuit into Clifford gates and $T$-gates, one can implement each $T$-gate by applying a $e^{i Z \pi/4}$-gate or no gate both with probability $1/2$, which is a $S$-gate times a global phase. Hence, because of the attenuation factor $\lambda=\cos(\pi/8)$ for each replaced $T$-gate, this enables one to sample classically from a circuit with $t$ many $T$-gates in a time $e^{2|\log(\cos \tfrac{\pi}{8})|t}$ multiplied by the simulation time of the resulting Clifford circuits. The exponent matches the currently best known exponent \cite{bravyi2016improved}.

We thank David Zsolt Manrique for helpful comments on the draft. E.G. acknowledges support by the Bavarian Ministry of Economic Affairs, Regional Development and Energy (StMWi) under project Bench-QC (DIK0425/01).

\bibliography{main}

\newpage

\onecolumngrid



\section{-SUPPLEMENTAL MATERIAL-}

\section{Optimal number of shots per circuit.}
In our algorithm, an expectation value is expressed as an average over random circuits. Each of these random circuits is then implemented on a quantum computer with a certain number of shots. Let us denote $S$ the number of shots per circuit and $N/S$ the number of circuits. At fixed $N$, we would like to find the value of $S$ that minimizes the variance on the result, taking into account both the shot noise and the statistical error due to the random circuits.

Given a random circuit $U$ drawn from the Poisson process, we denote $-1\leq m_U\leq 1$ the real part of the complex amplitude that we measure with our algorithm. Given the real part of a complex amplitude $-1\leq m\leq 1$, we denote $x_m$ the Bernoulli random variable that takes value $+1$ with probability $\frac{1+m}{2}$ and $-1$ with probability $\frac{1-m}{2}$. We  denote $\mathbb{E}$ the expectation value with respect to the random circuit $U$, and $\langle \rangle$ the expectation value with respect to the shot noise. The exact value of the real part of the complex amplitude that we measure after averaging over random circuits and shots is
\begin{equation}
    m=\mathbb{E}[\langle x_{m_U}\rangle]\,.
\end{equation}
With $S$ shots to evaluate $m_U$ and $N/S$ different circuits $U$, the estimated value $m_{\rm est}$ of $m$ is
\begin{equation}
    m_{\rm est}=\frac{1}{N/S}\sum_{p=1}^{N/S} \frac{1}{S}\sum_{i=1}^S x_{m_{U_p}}^{(i)}\,,
\end{equation}
where the $U_p$'s are independent circuits drawn from the Poisson process, and where the $x_{m_{U_p}}^{(i)}$'s are independent Bernoulli random variables with parameter $\frac{1+m_{U_p}}{2}$. Let us compute the variance of this quantity. We have
\begin{equation}
    \begin{aligned}
        m_{\rm est}^2=&\frac{1}{N^2}\sum_{i,j,p,q}x^{(i)}_{m_{U_p}}x^{(j)}_{m_{U_q}}\\
        =&\frac{1}{N}+\frac{1}{N^2}\sum_{\substack{i,j,p,q\\ i\neq j\text{ or }p\neq q}}x^{(i)}_{m_{U_p}}x^{(j)}_{m_{U_q}}\,,
    \end{aligned}
\end{equation}
where we used that $(x^{(i)}_{m_{U_p}})^2=1$. Then, averaging over the shots
\begin{equation}
\begin{aligned}
     \langle m_{\rm est}^2\rangle&=\frac{1}{N}+\frac{1}{N^2}\sum_{\substack{i,j,p,q\\ i\neq j\text{ or }p\neq q}}m_{U_p}m_{U_q}\\
    &=\frac{1}{N}+\frac{S^2}{N^2}\left(\sum_{p}m_{U_p}\right)^2-\frac{S}{N^2}\sum_{p}m_{U_p}^2\,.
\end{aligned}
\end{equation}
In the case where we have only one circuit $U$, i.e. $S=N$, we have $ \langle m_{\rm est}^2\rangle- \langle m_{\rm est}\rangle^2=\frac{1-m_U^2}{N}$, which is the usual variance of a Bernoulli random variable taking values $\pm 1$ with mean $m_U$. In the general case, after averaging over the circuits we have
\begin{equation}
    \mathbb{E}[\langle m_{\rm est}^2\rangle]=\frac{1}{N}+\frac{S^2}{N^2}\left(\frac{N^2}{S^2}\mathbb{E}[m_U]^2+\frac{N}{S}\mathbb{E}[m_U^2]-\frac{N}{S}\mathbb{E}[m_U]^2\right)-\frac{1}{N}\mathbb{E}[m_U^2]\,.
\end{equation}
Denoting $v=\mathbb{E}[m_U^2]-\mathbb{E}[m_U]^2$ the variance over the circuits, it yields
\begin{equation}
    \mathbb{E}[\langle m_{\rm est}^2\rangle]-\mathbb{E}[\langle m_{\rm est}\rangle]^2=\frac{1-v}{N}+\frac{S}{N}v\,.
\end{equation}
Since $v\geq 0$, we obtain that at fixed total number of shots $N$, doing only one shot $S=1$ per circuit minimizes the total variance.

\section{Optimal gate angles.}

When measuring an observable $\mathcal{M}$ with our algorithm with gate angles $\tau_n$, in absence of noise, the total number of rotations to perform to reach a precision $\epsilon$ on the result is $R(\{\tau_n\})/\epsilon^2$, where
\begin{equation}
    R(\{\tau_n\})=2t\sum_{n=1}^N \frac{|c_n|}{\sin\tau_n}\exp\left(4t\sum_{n=1}^N|c_n|\tan(\tau_n/2)\right)\,.
\end{equation}
Writing $\partial_{\tau_n}R=0$ at the optimal gate angle $\tau_n^*$, we find that it satisfies the equation
\begin{equation}
    \frac{2t}{\cos^2(\tau_n^*/2)}\sum_{p=1}^N \frac{|c_p|}{\sin\tau_p^*}=\frac{\cos\tau_n^*}{\sin^2\tau_n^*}\,.
\end{equation}
The left-hand side is bounded from below by $2t\mu$. Hence at large $t$, we necessarily have $\tau_n\to 0$ for all $n$ to make the right-hand side go to $\infty$. In that limit we have
\begin{equation}
    2t\sum_{p=1}^N \frac{|c_p|}{\tau_p^*}\sim (\tau_n^{*})^{-2}\,.
\end{equation}
The left-hand side is independent of $n$, so the leading behaviour of $\tau_n^*$ when $t\to\infty$ is independent of $n$. Denoting this leading behaviour $\tau^*$, we thus have $2t\mu (\tau^*)^{-1}=(\tau^{*})^{-2}$, and so
\begin{equation}
    \tau_n^*=\frac{1}{2t\mu}
\end{equation}
Let us now consider the noisy case where each rotation $e^{i\tau_nO_n}$ comes with an attenuation factor $e^{-r_n}$. The damping due to noise per circuit is thus $\exp\left(-2t\sum_{n=1}^N \frac{|c_n|r_n}{\sin\tau_n}\right)$. The total number of rotations to reach a precision $\epsilon$ is thus $R(\{\tau_n\})/\epsilon^2$, with now
\begin{equation}
     R(\{\tau_n\})=2t\sum_{n=1}^N \frac{|c_n|}{\sin\tau_n}\exp\left(4t\sum_{n=1}^N|c_n|\tan(\tau_n/2)+4t\sum_{n=1}^N \frac{|c_n|r_n}{\sin\tau_n}\right)\,.
\end{equation}
Writing again $\partial_{\tau_n}R=0$ at the optimal gate angle $\tau_n^*$, we find that it satisfies the equation
\begin{equation}
   2t \left(\frac{1}{\cos^2(\tau_n^*/2)}-\frac{2r_n \cos\tau_n^*}{\sin^2 \tau_n^*}\right)\sum_{p=1}^N \frac{|c_p|}{\sin\tau_p^*}=\frac{\cos\tau_n^*}{\sin^2\tau_n^*}\,.
\end{equation}
Let us assume that when $t\to\infty$, the quantity $Q\equiv\frac{1}{\cos^2(\tau_n^*/2)}-\frac{2r_n \cos\tau_n^*}{\sin^2 \tau_n^*}$ does not go to $0$. Then the same reasoning as above would apply and we would have $\tau_n^*\to 0$. But then we would have $Q\to-\infty$, and so the right-hand side would become negative, which cannot be. Hence we must have $Q\to 0$. This yields an equation for $\tau_n^*$ when $t\to\infty$
\begin{equation}
    \frac{1}{\cos^2(\tau_n^*/2)}=\frac{2r_n \cos\tau_n^*}{\sin^2 \tau_n^*}\,.
\end{equation}
Assuming $r_n$ small, this is
\begin{equation}
    \tau_n^*=\sqrt{2r_n}\,.
\end{equation}

\end{document}